\documentclass{iopart}
\usepackage{iopams}
\usepackage[scriptsize,bf]{caption}
\usepackage{xspace}
\eqnobysec
\newcommand{\const}{\mathrm{const.}}

\renewcommand{\br}[1]{\left(#1\right)}
\newcommand{\ud}[1]{\mathrm{d}#1}
\renewcommand{\exp}[1]{e^{#1}}
\newcommand{\sq}[1]{\left[#1\right]}

\newcommand{\sgn}{\mathrm{sgn}}
\newcommand{\abs}[1]{\left|#1\right|}

\renewcommand{\vec}[1]{\boldsymbol{#1}}

\newcommand{\dd}[1]{\dot{#1}\dot{#1}}
\newcommand{\ie}{\textit{i.e.}\xspace}

\newcommand{\eqref}[1]{\eref{#1}}
\newcommand{\cp}{$\mathbb{R}^3\times\mathbb{S}^2$\xspace}

\newcommand{\dof}{{d.o.f}\xspace}
\newcommand{\cmf}{\textit{CM} frame\xspace}

\begin{document}

\title[\underline{{\L}ukasz Bratek,  \ \ \ \ \   Fundamental
Relativistic Rotator. Hessian singularity\dots \ \ \ (preprint)}
]{Fundamental relativistic rotator: Hessian singularity
and\\ the issue of the minimal interaction with
electromagnetic field}

\author{{\L}ukasz Bratek}
\address{Henryk Niewodnicza{\'n}ski Institute of Nuclear Physics, \\
Polish Academy of Sciences, Radzikowskego 152, PL-31342 Krak{\'o}w,
Poland}
\ead{lukasz.bratek@ifj.edu.pl}

\begin{abstract}\\
There are two relativistic rotators with Casimir invariants of the Poincar\'{e}
group being fixed parameters. The particular models of spinning particles
were studied in the past both at the classical and quantum level. Recently, a
minimal interaction with electromagnetic field has been considered. We show
that the dynamical systems can be uniquely singled out from among other
relativistic rotators by the unphysical requirement that the Hessian referring to
the physical degrees of freedom should be singular. Closely related is the fact
that the equations of free motion are not independent, making the evolution
indeterminate. We show that the Hessian singularity cannot be removed by
the minimal interaction with the electromagnetic field. By making use of
a nontrivial Hessian null space, we show that a single constraint appears
in the external field for consistency of the equations of motion with the
Hessian singularity. The constraint imposes unphysical limitation on the initial
conditions and admissible motions. We discuss the mechanism of appearance
of unique solutions in external fields on an example of motion in the uniform
magnetic field. We give a simple model to illustrate that similarly constrained
evolution cannot be determinate in arbitrary fields.
\end{abstract}
\medskip
\hrule\medskip
\noindent
\textbf{The definitive version is available at \\ \texttt{http://iopscience.iop.org/1751-8121/44/19/195204/}}
\hrule

\section{Introduction}

The aim of this work is to elucidate at the Lagrangian level some unexpected indeterminacy in
the motion of a geometric spinning particle model. The model was originally proposed in \cite{bib:segal} and later rediscovered in quite a different context as the fundamental relativistic rotator \cite{bib:astar1}. We shall identify the cause of this indeterminacy and show that in the presence of external
fields a constraint appears imposing unphysical limitations on the motion and the freedom
in choosing the initial conditions. To understand better this singular behavior we found it instructive to contrast it with unique motion of other rotators from the family of relativistic
rotators defined in \cite{bib:astar1} both in the presence and absence of interactions. First, we recall what
relativistic rotators are and discuss the distinguished role fundamental rotators play among
them.

\subsection{Relativistic rotators}

The underlying motivation behind Staruszkiewicz paper \cite{bib:astar1}
was to design a mathematical mechanism suitable for an ideal classical
clock. This construction employed the notion of a rigid body of Hanson
and Regge \cite{bib:hanson} -- a rotating tetrad assigned to a
worldline. A rotating tetrad can be realized as a continuous action of
a proper ortochronous Lorentz operator on some initial tetrad. As
observed in \cite{bib:astar1}, such motion can be equivalently
described by specifying the motion of three distinct null directions.
Recall, that a cross-ratio of four independent null directions is an
invariant of  homographic transformations (being in 2-to-1
correspondence with the proper ortochronous Lorentz transformations
\cite{bib:penrose}). Hence, the motion of the fourth null direction
forming the tetrad can be uniquely determined from the  value of the
cross-ratio. It is  evident that the number of degrees of freedom
(\dof) is 9 (3 for position and 6 defining three null directions). A
rotator in Newtonian physics has only five \dof. Accordingly, a
relativistic counterpart of such a rotator is a dynamical system
consisting of position and a single null direction. This particular
example of a rigid body  is called a \textit{relativistic rotator}
\cite{bib:astar1}.

The canonical momenta of a rotating tetrad have too many \dof. This
arbitrariness is characteristic of the relativistic theory of spinning
bodies \cite{bib:dixon}. To make the equations of motion determinate,
it is customary to impose supplementary conditions reducing the number
of \dof  to the three rotational ones in the rest frame (the same as for
the Eulerian rigid body). But there are interesting exceptions. Already
a dynamical system described by a worldline and a single spinor of
fixed magnitude has the correct number of \dof: three for position, two for
the spinor's direction and one for the spinor's phase associated with
a rotation about the spinor's direction. A general form of the
Lagrangian for a single spinor can be found in \cite{bib:bratek3}. In
this framework, the Lagrangian of relativistic rotators can be
alternatively arrived at by neglecting the spinor's magnitude and its
phase. Then one is left with the following Hamilton's action:
\begin{equation}\label{eq:act_f} \fl S=-m\int\ud{\tau}
\sqrt{\dd{x}}f(Q),\quad Q\equiv{-\ell^2\frac{\dd{k}}{\br{k\dot{x}}^2}},
\quad kk=0, \quad f'(Q)\ne0,\quad f(Q)>0,\end{equation} the same as
originally defined in \cite{bib:astar1}. Here, a dot denotes
differentiation with respect to an arbitrary worldline parameter
$\tau$. The physical configuration space is identical to \cp. A point
in $\mathbb{R}^3$ represents the actual position whereas a point on the
unit sphere $\mathbb{S}^2$  the actual direction in the physical space.
By the symmetry argument, neither a particular initial position nor the direction
can be distinguished. The same concerns the associated initial
velocities. The additional two auxiliary \dof present in the Lagrangian
are gauge \dof of the reparametrization and projection invariance,
namely, the worldline parameter and the scale of the null vector.  They
do not alter the physical state and can be arbitrary functions.

To reduce the variety of Lagrangians possible in \eqref{eq:act_f}, it
is  postulated that both Casimir invariants of the Poincar\'{e} group
should be fixed parameters rather than constants of motion
\cite{bib:astar1}. This provides us with two distinct conditions  which
we shall call \textit{fundamental conditions}:
\begin{equation}\label{eq:fundam_condit}PP=m^2,\qquad
WW=-\frac{1}{4}m^4\ell^2.\end{equation} Here, $W^{\mu}$ is the
Pauli-Luba\'{n}ski spin-pseudovector
$W^{\mu}=-\frac{1}{2}\epsilon^{\mu\alpha\beta\gamma}M_{\alpha\beta}P_{\gamma}$,
where $P_{\mu}$ and $M_{\mu\nu}$ are Noether constants of motion. Owing
to property \eqref{eq:fundam_condit}, a classical mechanical system
can be identified by specifying its invariant mass and spin. This
feature can be regarded at the classical level as the counterpart of
irreducibility characteristic of relativistic quantum states. By this
analogy, relativistic dynamical systems for which
\eqref{eq:fundam_condit} hold are called \textit{fundamental dynamical
systems} and the other are called \textit{phenomenological}
\cite{bib:astar1}. It is a simple matter to demonstrate that there are
only two relativistic rotators that are fundamental. Their trajectories
are extremals of the Hamilton's action (cf \ref{app:fr}):
\begin{equation}\label{eq:action}S=-m\int{\ud{\tau}\sqrt{\dot{x}\dot{x}}
\sqrt{1\pm\sqrt{-\ell^2\frac{\dot{k}\dot{k}}{\br{k\dot{x}}^2}}}}.\end{equation}
The existence of the two rotators is remarkable since there is no
apparent reason for two differential equations for a single function $f$,
originating from two distinct notions of mass and that of spin, to  have
a common solution. Observe the exceptional fact that for the action of
the form \eqref{eq:act_f} the requirement that the Casimir mass is a
fixed parameter implies that the Casimir spin is also fixed, and vice
versa (there is no such implication for more complicated dynamical
systems, e.g. cf \cite{bib:bratek2}).

As it happened many times in the history of science, quite unrelated
motivations and various ways of thinking make people to come up,
independently, with similar ideas. It should be mentioned that
Lagrangian \eqref{eq:action} was discovered before Staruszkiewicz by
Kuzenko, Lyakhovich and Segal \cite{bib:segal}. To find a unique
Lagrangian for their geometric model of a spinning particle, they also
imposed the condition of constant mass and spin, which they called
\textit{strong conservation}.

Staruszkiewicz suggested that a dynamical system like \eqref{eq:action}
is an ideal mathematical clock that could be used in studying some
difficult and not well-understood problems in special and general
theory of relativity \cite{bib:astar1}. The clocking mechanism is
simple. The image of the spatial direction $W^{\mu}$ is represented by
a fixed circle on the Riemann sphere of complex numbers, whereas the
image of the null direction $k^{\mu}$ is represented by a point moving
about that circle. The cycles of the clock are measured by means of
counting the number of times the phase assigned to the circular
motion has been increased by $2\pi$. Accordingly, the images of
$W^{\mu}$ and $k^{\mu}$  can be regarded  as the clock's dial and the
clock's hand, respectively.

\subsection{Outline of this work}

It is important to exclude \textit{irregular} Lagrangians on \cp. To
find such Lagrangians one simply has to require that a corresponding
Hessian determinant must be vanishing. This condition will give rise to
some second-order differential equation for function $f$ present in
\eqref{eq:act_f}. Surprisingly, it will also lead to the action functional \eqref{eq:action}. In this case only four equations of
motion are independent. To identify the indeterminate degree of freedom
we construct in a covariant manner a parametric description of
the most general solution.

We shall also examine the motion of relativistic rotators in the
presence of the electromagnetic field. This is done not without a reason.
The customary form of the interaction assumed for charged particles
cannot remove the Hessian singularity and a \textit{Hessian constraint}
appears closely linked with this singularity. It constrains the
evolution on \cp and limits the freedom in choosing the initial data,
breaking the symmetry of \cp and of the tangent space, even in the
limit of infinitely weak fields.  In finding constraints of this kind
the concept of the null space of a singular Hessian will show up
useful. With this constraint there is no reason to expect unique
solutions for arbitrary fields. However, there are unique solutions
possible when stronger constraints compatible with the Hessian
constraint are assumed, in which case variations in the direction of
the Hessian null space are excluded. All these observations will lead
us to the conclusion that relativistic rotators with identically fixed
mass and spin are unphysical. In contrast, for regular Lagrangians on
\cp, the motion will turn out unique and not constrained, both in the
presence and absence of fields.

Following the Staruszkiewicz paper \cite{bib:astar1}, there has been a
growing interest in the fundamental relativistic rotator
\cite{bib:schaefer,bib:hind}. However, the central deficiency of the
Lagrangian, showing up in the Hessian singularity and the associated
indeterminacy of the equations of motion on \cp, has been unnoticed
(except for in papers \cite{bib:bratek3} and \cite{bib:bratek2}, where
this singularity was already alluded to in the context of more general models, but compare also the original unpublished paper
\cite{bib:bratek1} where the central results were found more than 2
years ago already and become the basis for this paper.

\subsection{Notations and conventions} $ab$ stands for the scalar
product of $a$ and $b$:
$ab\equiv{}a^0b^0-a^1b^1-a^2b^2-a^3b^3=a^0b^0-\vec{a}\vec{b}$;
$\epsilon^{\mu\nu\alpha\beta}$ is a completely antisymmetric
pseudotensor for which $\epsilon^{0123}=1$. This metric signature
requires appropriate sign in the definition of canonical momenta, in
particular, $p_{\mu}=-\partial_{\dot{x}^{\mu}}L$.

\section{The Cauchy problem for the fundamental relativistic
rotator}\label{sec:cauchy}

We started off with a relativistically invariant action integral
\eqref{eq:action}.
To enable a manifestly covariant description obeying the
constraint $kk=0$ we need seven \dof, whereas only five \dof uniquely
define the physical state of a rotator. It would be aesthetically
compelling to retain a covariant description, but we want to focus on
another aspect of the model in which the auxiliary \dof are
unnecessary. We can eliminate the \dof at the cost of
loosing covariance of the description at the level of the equations of
motion, but we know our description is still relativistic in content.

From now on we shall be assuming that all of the auxiliary \dof have
been already eliminated from the action. That means that we are using a
quintuple of coordinates on the configuration space \cp and a fixed
time variable. It is convenient to introduce the natural coordinates on
\cp, $q\equiv\{q^1,q^2,q^3,q^4,q^5\}=\{x^1,x^2,x^2,\theta,\phi\}$ in
the gauge $\tau\equiv x^0$ and $k^0\equiv1$. In effect we obtain some
Lagrangian defined on \cp with the time variable being set to be the
Newtonian time. Let such reduced Lagrangian be denoted by
$L_N(v,q)$, where $q$ stands for the set of \dof on \cp,  and
$v\equiv\dot{q}$ for the associated velocities. Now, we can apply the
elementary Lagrangian or Hamiltonian formalism avoiding the issue of
constraints that would be necessary had we retained the auxiliary
variables. Yet there is one reservation left. The necessary condition
for the elementary Hamiltonian formulation is that the set of equations
$p(v,q)=\frac{\partial{}L_N}{\partial{}v}(q,v)$ defining momenta $p$
conjugated to $q$ be a diffeomorphism of the space of momenta $p$ and
that of velocities $v$ for all $q$. In other words, the set of
equations should be uniquely solvable for the velocities, $v=v(q,p)$.
It will be possible, provided the Hessian determinant is nonzero:
\begin{equation}\label{eq:necessary_condition}\det
\sq{\frac{\partial{}^2L_N}{\partial{}\dot{q}^i\partial\dot{q}^j}}\ne0.
\end{equation} Otherwise, the Legendre transformation leading from the
Lagrangian $L_N(v,q)$ to the Hamiltonian would not be well defined in
terms of $q$ and $p$. Moreover, for the Lagrangian equations can be
recast in a general form
$$\frac{\partial{}^2L_N}{\partial{}\dot{q}^i\partial\dot{q}^j}\,
\ddot{q}^j=Z_i(q,\dot{q},t),$$ with some function $Z$, condition
\eqref{eq:necessary_condition} is also necessary  for a unique
dependence of accelerations $\ddot{q}$ on positions $q$ and velocities
$\dot{q}$. The vanishing of the Hessian determinant would not only mean
that  $\ddot{q}$ would be non-unique, but also that the Lagrangian
equations could not be reduced to the canonical form  $\dot{y}=F(y,t)$,
where $y=\{q,\dot{q}\}$, for which the textbook results on the
existence and uniqueness are known for solutions of ordinary
differential equations. In particular, the Lagrangian equations could
not be solved directly by means of the Picard method or a numerical step by
step integration. A similar viewpoint on the necessity of
\eqref{eq:necessary_condition} for internal coordinates is presented
for Lagrangian systems in \cite{bib:gitman}. With a singular Lagrangian
there would be a gauge freedom or constraints in the system. Both
situations would be physically unacceptable for rotators with
Lagrangians already expressed in terms of the internal coordinates on
\cp and a fixed time variable.

\subsection{\label{sec:characterisation}Characterization of fundamental
relativistic rotators by Hessian singularity}

In the natural coordinates on \cp the
Lagrangian $L_N$ corresponding to \eqref{eq:act_f} reads
\begin{equation}\label{eq:dynamical_lagrangian} \fl
L_N=-m\sqrt{1-\dot{\vec{x}}\dot{\vec{x}}}\,f(Q),\qquad
Q=\ell^2\frac{\dot{\vec{n}}\dot{\vec{n}}}{\br{1-\vec{n}\dot{\vec{x}}}^2}
, \quad f'(Q)\ne0,\quad f(Q)>0,\end{equation} where
$\vec{x}=(x^1,x^2,x^2)$,
$\vec{n}=(\sin{\theta}\cos{\phi},\sin{\theta}\sin{\phi},\cos{\theta})$
and $|\dot{\vec{n}}|=\sqrt{\dot{\theta}^2+\dot{\phi}^2\sin^2{\theta}}$.

By a \textit{Hessian} assigned to the Lagrangian $L_N$ we mean a square
(symmetric) matrix of second partial derivatives of $L_N$ with respect to
velocities $\{\dot{x}^1,\dot{x}^2,\dot{x}^2,\dot{\theta},\dot{\phi}\}$.
The Hessian determinant was calculated in \ref{sec:hessian_calculation}
and reads \begin{equation}\label{eq:hessian} \det{
\sq{\frac{\partial{}^2L_N}{\partial{}\dot{q}^i\partial\dot{q}^j}}}\propto
f(Q)^3f'(Q)^2 \br{1+2\,Q\br{
\frac{f'(Q)}{f(Q)}+\frac{f''(Q)}{f'(Q)}}}.\end{equation} Here is shown
only the physically relevant universal part of the Hessian determinant.
It is an $f$-dependent Lorentz scalar uniquely determined by the
structure of the model.  The omitted proportionality factor is not of
interest and may change depending on the particular coordinates used on
\cp.

It follows that the only nontrivial function $f(Q)$ for which the
Hessian determinant is identically zero is
$f(Q)=c_1\sqrt{1+c_2\sqrt{Q}}$. Here, $c_1$ and $c_2$ are some
integration constants and they can be absorbed by the dimensional
parameters $m$ and $\ell$ of the model. This gives us
$$f(Q)=\sqrt{1\pm\sqrt{Q}}.$$  Hence, there are only two physically
distinct Lagrangians with singular Hessian. Again, we arrive at
action \eqref{eq:action} obtained in \ref{app:fr} from fundamental
conditions \eqref{eq:fundam_condit}.

\subsection{\label{sec:gensol}The Hessian null space and the free
motion of the fundamental relativistic rotator}

We have seen in section \ref{sec:characterisation} that the Lagrangian
of the fundamental relativistic rotator is irregular on \cp.  Now let
us see how this defect reflects in the free motion of the rotator. In
our parametrization used on \cp the Lagrangian in \eqref{eq:action}
attains the form \begin{equation}\label{eq:dynamical_lagrangian}
L_N=-m\sqrt{1-\dot{\vec{x}}\dot{\vec{x}}}
\sqrt{1+\ell{\frac{|\dot{\vec{n}}|}{|1-\vec{n}\dot{\vec{x}}|}}}.\end{equation}
For concreteness we have assumed $f(Q)=\sqrt{1+\sqrt{Q}}$  (the results
are quite analogous for $f(Q)=\sqrt{1-\sqrt{Q}}$).

A singular Hessian has a nontrivial null space. In
\ref{sec:null_space_calc} it was shown that the Hessian null space for
\eqref{eq:dynamical_lagrangian} is spanned by a single vector. We call
it the \textit{kernel vector} and denote by $w$. The vector $w$ is
represented on \cp as  (see equation \eqref{eq:nullvector})
\begin{equation}\fl\label{eq:null_vector}w=\frac{\ell}{2}\abs{\dot{\vec{n}}}
\br{\vec{n}-\dot{\vec{x}}}\partial_{\vec{x}}\oplus\varrho\,
({\dot{\theta}}\,\partial_{\theta}+{\dot{\phi}}\,\partial_{\phi}),\quad
\varrho=
\frac{(1-\vec{n}\dot{\vec{x}})^2+(\vec{n}-\vec{\dot{x}})^2\frac{\ell}{2}
\abs{\dot{\vec{n}}}}{ {1-\dot{\vec{x}}\dot{\vec{x}}}}. \end{equation}
In general, to every nontrivial kernel vector
$w_{(a)}=w_{(a)}(q,\dot{q},t)$ of an irregular Lagrangian $L$ a single Hessian constraint is assigned
 involving positions $q$ and velocities
$\dot{q}$: $$\frac{\delta L}{\delta q^i}=0\quad\Rightarrow\quad
0=w_{(a)}^i\frac{\delta L}{\delta q^i} =
w_{(a)}^i\br{\frac{\ud}{\ud{t}}\!\!\sq{\frac{\partial{}L}{\partial{}\dot{q}^i}}
-\frac{\partial{}L}{\partial{}{q}^i}}=w^i_{(a)}(q,\dot{q},t)\,Z_i(q,\dot{q},t),$$
where we have used the fact that $w^i\frac{\partial^2L}{\partial
\dot{q}^i\partial \dot{q}^j}\,\ddot{q}^j=0$. The number of Hessian
constraints is  equal to the dimension of the Hessian null space.
Surprisingly, the single constraint for the kernel vector
\eqref{eq:null_vector} is trivial since the identity
\begin{equation}\label{eq:constraint_free_explicit} w^i\frac{\delta
L_N}{\delta q^i}= w^i\frac{\partial
\Gamma}{\partial{\dot{q}^i}}-2\varrho\,\Gamma\equiv0, \qquad
\Gamma=\dot{\theta}\frac{\partial L_N}{\partial
\theta}+\dot{\phi}\frac{\partial L_N}{\partial \phi} \end{equation}
holds for the Lagrangian \eqref{eq:dynamical_lagrangian} independently
of whether the equations of motion on \cp are satisfied or not. The
identity \eqref{eq:constraint_free_explicit} states that the system of
Lagrangian equations for the Lagrangian \eqref{eq:dynamical_lagrangian}
is linearly dependent and therefore under-determined. Only four among five
equations are independent. As a consequence, there will be a single
arbitrary function of the time present in the general solution.

 As follows from the calculation in \ref{sec:solution}, the most
 general solution to the equations of free motion of the fundamental
 relativistic rotator in the gauge $p\dot{x}=m$ and $pk=m$, has the
 following covariant parametric description:
 \begin{equation}\label{eq:gensol}\fl
 x^{\mu}(t)=\frac{P^{\mu}}{m}\,t+\frac{\ell}{2}\,r^{\mu}(t)+x^{\mu}(0),\quad
 \mathrm{and}\quad
 k^{\mu}\br{t}=\frac{P^{\mu}}{m}+\frac{\dot{r}^{\mu}(t)}{{\sqrt{-\dot{r}(t)\dot{r}(t)}}},
 \end{equation} where $$r^{\mu}(t)=N^{\mu}\sin{\phi(t)}+
 \frac{\epsilon^{\mu\nu\alpha\beta}N_{\nu}W_{\alpha}P_{\beta}}{
 \frac{1}{2}m^3\ell}\cos{\phi(t)}.$$ Constant vectors $P^{\mu}$,
 $W^{\mu}$ and $N^{\mu}$ satisfy the conditions \[\fl PP=m^2,\quad
 WW=-\frac{1}{4}m^4\ell^2,\qquad WP=0,\quad NN=-1,\quad NW=0,\quad
 NP=0.\] $P^{\mu}$ is the (conserved) momentum defining the time axis
 of the center of momentum frame (\textit{CM}), $t$ is the proper time in that frame,
 and $W^{\mu}$ is the (conserved) intrinsic angular momentum. Being the
 elliptic angle of a Lorentz operator, $\phi$ is a Lorentz scalar (see
 \ref{sec:solution}). It describes the angular position of the image of
 the null direction $k^{\mu}(t)$ on a large circle lying  on the
 Riemann sphere in the \cmf. The velocity of rotation relative to
 \cmf can be determined from the hyperbolic angle $\Psi$ between
 timelike vectors $P^{\mu}$ and $\dot{x}^{\mu}$,
 $\dot{x}P=|\dot{x}||P|\cosh{\Psi}$. Since
 $\dot{x}=\frac{P}{m}+\frac{\ell}{2}\,\dot{r}$ by \eqref{eq:gensol},
 we get  $|\dot{x}|\cosh{\Psi}=\frac{\dot{x}P}{m}=1$.   In addition,
 $\dot{x}\dot{x}=1+\frac{\ell^2}{4}{\,\dot{r}^2}$ and
 $\dot{r}^2=-\dot{\phi}^2$; hence,
 $|\dot{x}|\sinh{\Psi}=\frac{\ell}{2}|\,\dot{\phi}|$, which gives us
 the actual frequency of rotation $\dot{\phi}(t)$ in the \cmf
 \begin{equation}\label{eq:frequency}|\dot{\phi}(t)|=\frac{2}{\ell}\,\tanh{\Psi},
 \end{equation} numerically equal to a Lorentz scalar $\Psi$
 (rapidity). The Hamilton action \eqref{eq:action} evaluated for the
 general solution \eqref{eq:gensol} reads
 \[S(t)=S(0)-m\,t-m\frac{\ell}{2}\int|\dot{\phi}(t)|\,\ud{t}.\] The first
 term is the ordinary contribution from the inertial motion of the
 \cmf, growing linearly with  $t$. The second term contains the angle
 variable $\phi$ conjugate to the spin $\frac{1}{2}m\ell$. It does not
 grow linearly with $t$; it is left completely undetermined and  can be
 \textit{arbitrary} function of the time, such that $
 \frac{\ell}{2}\,|{\dot{\phi}(t)}|<1$. This upper bound by the velocity
 of light appears naturally in the solution since $\tanh{\Psi}<1$. The
 sign of ${\dot{\phi}(t)}$ should be constant during motion. Otherwise,
 there would be discontinuities in  $k^{\mu}(t)$, whereas the evolution
 of a dynamical system should be continuous (the part of $k^{\mu}$
 orthogonal to $P^{\mu}$ inverts its direction whenever $\dot{\phi}(t)$
 passes through $0$). This limitation, however, is not naturally
 implied by the solution and must be imposed by hands, which is another
 trace of defectiveness of the Lagrangian
 \eqref{eq:dynamical_lagrangian}. Had the rotator been a well-behaved
 dynamical system with non-singular Hessian, the ${{\phi}(t)}$ would
 be a linear function of $t$ and the flops in the direction of
 $k^{\mu}$ would be impossible. The point $\dot{\phi}\br{t}=0$ is a
 singularity of the equations of motion. It separates two qualitatively
 different regimes of the motion: the inertial free motion on a
 straight line with $\dot{{\phi}}(t)\equiv0$ and the free motion on a
 circle of fixed radius $\ell/2$ (as perceived in the \cmf).

\section{The Hessian singularity and the issue of interactions  with
external fields.}

The irregular Lagrangian \eqref{eq:action} could be supplemented  with
an additive interaction term in the hope of removing the free motion
indeterminacy. However, even when the interaction was suitable for
removing the Hessian singularity for nonzero fields, the problem of
indeterminate free motion would reoccur after switching off the fields.
The situation gets even worse when the Hessian singularity persists in
the fields. This can be seen in the example of electrically charged
fundamental relativistic rotator minimally interacting with
electromagnetism. Even though the model is unphysical, considering it
is still pedagogically instructive when contrasted with the motion of
relativistic rotators with regular Lagrangian on \cp for which such
form of interaction is quite permissible.

\subsection{\label{sec:constr_from_eq}Relativistic rotators minimally
coupled to external electromagnetic field}
If a relativistic rotator was to be treated as a structureless point
particle with electric charge $e$, it could be minimally coupled with
the electromagnetic field. In principle, the internal structure of the
rotator should also be taken into account  (through non-minimal terms
involving gauge-invariant scalars such as
$F_{\mu\nu}\frac{k^{\mu}\dot{k}^{\nu}}{\br{k\dot{x}}^2}$,
$F_{\mu\nu}\frac{k^{\mu}\dot{x}^{\nu}}{k\dot{x}}$, etc.). The minimal
interaction gives us the action integral  of the form
\begin{equation}\fl\label{eq:action_f_EM}\widetilde{S}=S+S_{I}=-m\int\ud{\tau}
\sqrt{\dd{x}}\br{f\br{Q}+\frac{e}{m}
\frac{A\dot{x}}{\sqrt{\dd{x}}}},\quad\  f'(Q)\ne0,\quad f(Q)>0.
\end{equation}
The resulting evolution law for the kinematical part of momentum is
$\dot{p}_{\mu}=e F_{\mu\nu}\dot{x}^{\nu}$. When projected onto the
direction of $p^{\mu}$, it gives us the evolution law for
$p_{\mu}p^{\mu}$: \begin{equation}\label{eq:constr1} \frac{1}{2}
\frac{\ud}{\ud{\tau}} \br{\frac{p_{\mu}p^{\mu}}{m^2}}=\frac{e}{m^2}
F_{\mu\nu}p^{\mu}\dot{x}^{\nu}=-2\frac{e}{m} Q
f'(Q)\frac{|\dot{x}|}{k\dot{x}}F_{\mu\nu}k^{\mu}\dot{x}^{\nu}.\end{equation}
In general, the scalar $p_{\mu}p^{\mu}$  will be variable for
phenomenological rotators. For example, this would occur in constant
magnetic field for initial conditions chosen so as the velocity vector
$\dot{\vec{x}}$ was not coplanar with vectors $\vec{B}$ and $\vec{k}$.
In order to have invariable $p_{\mu}p^{\mu}$ during motion, one could
try to look for solutions obeying  the condition
$F_{\mu\nu}k^{\mu}\dot{x}^{\nu}=0$, or alternatively,
$F_{\mu\nu}k^{\mu}{p}^{\nu}=0$. For fundamental rotators, however,
$p_{\mu}p^{\mu}$ is by construction a fixed parameter. Then
$F_{\mu\nu}k^{\mu}\dot{x}^{\nu}=0$ is not a possibility but a
constraint, an integrability condition that must be imposed always for
all solutions for consistency with the equations of motion, if only one
wants $Q\not\equiv0$. A similar derivation of this constraint was
presented  in \cite{bib:schaefer}.

 Later, in section \ref{sec:hessianconstr}, we shall derive constraint
 $F_{\mu\nu}k^{\mu}\dot{x}^{\nu}=0$ for fundamental rotators
 rigorously based on non-triviality of the Hessian null space on \cp.
 But already the following observation shows that this constraint is
 closely related to the Hessian singularity and is absent when the
 Hessian is not singular. Namely, by making use of the identity
 $p_{\mu}p^{\mu}\equiv{}m^2f(Q)\br{f(Q)-4Qf'(Q)}$, equation
 \eqref{eq:constr1} can be written in the form involving a term with
 which we are familiar from the Hessian \eqref{eq:hessian}
\begin{equation}\label{eq:frequency_evolution}\br{1+2\,Q\br{\frac{f'(Q)}{f(Q)}+
\frac{f''(Q)}{f'(Q)}}}\frac{\ud{Q}}{\ud{s}}=
\frac{2Q}{f(Q)}\cdot\frac{e}{m}\frac{F_{\mu\nu}k^{\mu}\dot{x}^{\nu}}{k\dot{x}},
\end{equation}
where $\ud{s}=\sqrt{\dot{x}\dot{x}}\,\ud{\tau}$.
It follows that function $Q$ will  be constant for phenomenological
rotators when $F_{\mu\nu}k^{\mu}\dot{x}^{\nu}=0$. In that case, the
actual value of the rotation velocity relative to the instantaneous
\cmf, numerically equal to
$$\tanh\Psi=\frac{2\,Qf'(Q)}{f(Q)-2Qf'(Q)}, \qquad \mathrm{where}\
\cosh{\Psi}=\frac{p\,\dot{x}}{|p||\dot{x}|},$$
 will also be constant. In particular, this would be the case in free
 motion. For arbitrary field, equation \eqref{eq:constr1} can be viewed
 as the evolution law for the instantaneous frequency of rotation
 relative to the \cmf.
However, the situation is qualitatively different when the Hessian
determinant \eqref{eq:hessian} is  identically vanishing. In that case,
as we have already seen in section \ref{sec:gensol},
the frequency of rotation is not constant in free motion; it is
indeterminate  and can be an arbitrary function of the time (cf
\eqref{eq:frequency}). That is consistent  with
\eqref{eq:frequency_evolution}. Furthermore, in the presence of
the electromagnetic field, if one does not want to have always $Q=0$ as for
a point charge,  one has to impose on solutions the constraint
$F_{\mu\nu}k^{\mu}\dot{x}^{\nu}=0$, since then  the left-hand side in
equation \eqref{eq:frequency_evolution} vanishes identically. Then,
instead of the evolution law for $Q$ admitting all initial conditions
for rotators with regular Lagrangians on \cp, one is left with a
constraint both for the motion and for the initial conditions.

\subsection{\label{sec:hessianconstr}Hessian constraint in the presence
of the electromagnetic field}

In section \ref{sec:constr_from_eq} we derived the constraint
$F_{\mu\nu}k^{\mu}\dot{x}^{\nu}=0$ from the equations of motion for
rotators described by the free action part \eqref{eq:action}  and we
have seen from \eqref{eq:frequency_evolution} that the appearance of
the constraint must be related to the identical vanishing of the
Hessian determinant on \cp. Now, we shall derive the constraint in a
way directly referring to the algebraic properties of the singular
Hessian, by making use of its nontrivial null space.

To this end we may confine ourselves  to considering the particular
parametrization as in \eqref{eq:dynamical_lagrangian}  assuming the
$+$ sign  (for the $-$ sign the results will be analogous). We remind
that the fundamental relativistic rotator is not structureless;
however, similarly as in \eqref{eq:action_f_EM} we assume for
pedagogical reasons the minimal interaction with the electromagnetic
field. Then the Lagrangian \eqref{eq:dynamical_lagrangian} should be
supplemented with the interaction term $$L_{I}=-e
A\dot{x}=e\,\dot{\vec{x}}\vec{A}(\vec{x},t)-e\,\Phi(\vec{x},t),$$ where
$\Phi$ and $\vec{A}$ form  the electromagnetic potential. The Hessian
matrix \eqref{eq:hessian_matrix} is not altered by this interaction;
thus, the Hessian determinant is still vanishing and the kernel vector
$w$  remains the same as in \eqref{eq:null_vector}. By taking into
account identity \eqref{eq:constraint_free_explicit} which holds
for the free Lagrangian part, we have
$$\frac{\ell}{2}\abs{\dot{\vec{n}}}\br{\vec{n}-\dot{\vec{x}}}
\frac{\delta L_I}{\delta
\vec{x}}+\rho\underbrace{\br{\dot{\theta}\,\frac{\delta L_I}{\delta
\theta} +\dot{\phi}\,\frac{\delta L_I}{\delta \phi}}}_{=0}
=-e\frac{\ell}{2}\abs{\dot{\vec{n}}}
\br{\vec{n}-\dot{\vec{x}}}\br{\vec{E}+\dot{\vec{x}}\times\vec{H}}, $$
which leads to the following Hessian constraint when
$|\dot{\vec{n}}|\neq0$:
\begin{equation}\label{eq:constraint_vector_form}
\br{\vec{n}-\dot{\vec{x}}}\br{\vec{E}+\dot{\vec{x}}\times\vec{H}}=0
\quad \Longleftrightarrow \quad
\vec{n}\br{\vec{E}+\dot{\vec{x}}\times{\vec{H}}}=\dot{\vec{x}}\vec{E}.
\end{equation} Coming back to the covariant notation, we obtain
$F_{\mu\nu}k^{\mu}\dot{x}^{\nu}=0$ which is familiar from the previous
paragraph. In contrast, for a nonsingular Hessian on \cp the Hessian
null space is trivial and the same reasoning gives the identity $0=0$.
In that case any nontrivial linear combination of the equations of
motion would always result in some identities like
\eqref{eq:frequency_evolution}, which are not  constraints.

\subsection{\label{sec:magnetic} Occurrence of unique solutions in
external fields for  irregular Lagrangians}

Free motion of a dynamical system with singular Hessian can be
non-unique, described by arbitrary functions of the time. The
non-uniqueness we obtained for fundamental rotators is real, since on
the one hand $\Psi$, the rapidity, is an observable  and, on the other hand, it behaves as a
gauge variable. It must not be so for a well behaved
dynamical system whose physical state must be unique. When an
interaction term is added and the singularity is not removed, new
constraints may appear. We called them Hessian constraints. Although
constraints impose additional limitations on solutions, they are not
sufficient to render the motion unique for all field configurations and
initial conditions. A possible criterion to decide if a constraint
consistent with the Hessian constraint leads to a unique solution is to
check if it excludes variations of velocities in the null direction of
the Hessian.

\subsubsection{\label{sec:example}An example.} For the purpose of
illustration, it will be instructive to consider the following
non-relativistic model of a point particle. Let its intrinsic structure
be described in cylindrical coordinates $r$, $\phi$, $z$ by the
Lagrangian
\[\fl L_o=\case12{}m\br{\dot{r}^2+r^2\dot{\phi}^2+\dot{z}^2}+\case18{}m\ell^2\dot{\psi}^2-
\case12{}m\ell\,|\dot{\psi}|\br{\dot{r}
\cos\br{\psi-\phi}+r\dot{\phi}\sin\br{\psi-\phi}}.\]  The fourth
coordinate $\psi$ refers to some internal rotational degree of freedom
of the particle. Alternatively, one can regard this Lagrangian as
describing a structureless point particle living in a four-dimensional
hyper-cylinder $\mathbb{R}^3\times\mathbb{S}^1$ with the fourth dimension
$\psi$ being curled up into a tiny circle of radius $\case12\ell$, and
moving in a complicated velocity-dependent field. The Hessian is
singular. Its only kernel vector is
$w=r\cos\br{\psi-\phi}\partial_r+\sin\br{\psi-\phi}\partial_{\phi}+
\case{2r\epsilon}{\ell}\,\partial_{\psi}$,
$\epsilon=\sgn{(\dot{\psi})}$. The associated Hessian constraint is
$w^i\case{\delta L_o}{\delta q^{i}}\equiv0$. It vanishes identically
(not only for solutions). Hence, the equations of motion are not
independent, similarly as for the Lagrangian
\eqref{eq:dynamical_lagrangian} and the motion is non-unique. In
particular, there is a solution $r=\case{\ell}{2}$, $\phi=\nu(t)$,
$z=0$, $\psi=\nu(t)+\epsilon\,\case{\pi}{2}$,  with an arbitrary function
$\nu(t)$. Solutions with various $\nu$ and satisfying the same initial
conditions are indistinguishable based on the least action principle.

What the presence of interactions would change? In the field of the
electric type $V_E=Kz$, the Hessian constraint is still trivial since
$w^i\case{\delta V_E}{\delta q^{i}}=0$  and $\nu(t)$ is still
non-unique, with the only difference that the center of mass
accelerates  along the $z$ axis, $\ddot{z}=-\case{K}{m}$. In the field
of the magnetic type, $V=\case{1}{2}\tilde{K}r^2\dot{\phi}$, the
Hessian constraint is nontrivial for nonzero $\tilde{K}$
$r\dot{\phi}\cos\br{\psi-\phi}-\dot{r}\sin\br{\psi-\phi}=0$. Here, we
mention only some fixed frequency solutions that are unique: a)
$\phi=\omega t$, $\psi=\omega t-\case{\pi}{2}$,
$\omega=\case{\tilde{K}R}{m(R+\ell/2)}$ for $R>0$; b) $\phi=-\omega t$,
$\psi=-\omega t-\case{\pi}{2}$, $\omega=\case{\tilde{K}R}{m(\ell/2-R)}$
for $R<\ell/2$; and  c) $\phi=\omega t$, $\psi=\omega t+\case{\pi}{2}$,
$\omega=\case{\tilde{K}R}{m(R-\ell/2)}$ for $R>\ell/2$, where, for all
these solutions, $z=0$ and $r=R>0$ (all constants assumed positive).

It is clear from this example that the addition interactions cannot
restore uniqueness in all cases.  To make the motion always unique in
an arbitrary field, the Hessian must be nonsingular -- it would suffice to
change a little the $1/2$ factor  standing in front of the second term in
the Lagrangian  $L_o$ to achieve this  (then no Hessian constraint
would appear).

\subsubsection{\label{sec:magnetic_remark} Occurrence of unique
solutions for charged fundamental  rotator. Uniform magnetic field.} We
have seen above that  solutions may occasionally occur  to be unique
in the presence of external fields despite Hessian singularity. A
similar behavior should be expected for the fundamental rotator. An
example of such motion was found in the uniform magnetic field in
\cite{bib:schaefer} but the reason for its uniqueness was left
unexplained, however. There is also an apparent contradiction that
should be explained: these motions  are unique for the arbitrarily small
magnetic field, whereas  the free motion is not unique. Paradoxically,
the Hessian singularity will play a central role in understanding  this
uniqueness issue.

In free motion of the fundamental rotator,  vectors $\vec{n}$ and
$\dot{\vec{x}}$ are co-rotating in the \cmf, \ie
$\vec{n}\times\dot{\vec{x}}=0$, and the motion is circular in that
frame.  Co-rotation is also trivially  consistent with the condition
$\vec{H}\br{\vec{n}\times\dot{\vec{x}}}=0$  to which the Hessian
constraint \eqref{eq:constraint_vector_form} reduces in the presence of
pure magnetic field in some reference frame. Recall, that the motion of
a point charge is circular in the plane perpendicular to  $\vec{H}$,
has fixed rotational frequency and becomes inertial in the limit
$|\vec{H}|\to0$ when the frequency tends to $0$. By this analogy one
can expect the frequency to be fixed also for the fundamental rotator
in co-rotational circular motion  in the plane perpendicular to
$\vec{H}$.

To discuss the particular motions it will be useful to work in
cylindrical coordinates with the $z$ axis directed along $\vec{H}$:
$\vec{H}=[0,0,H]$, $H>0$. For co-rotational motion with frequency
$\dot{\phi}$ about a circle of fixed radius $R$,
$\dot{\vec{x}}=\epsilon\,R\,\dot{\phi}\,\vec{n}$, where
$\epsilon=\pm1$; hence, $\vec{H}\vec{n}=0$ since
$\dot{\vec{x}}\vec{H}=0$. Thus, the co-rotation ansatz is a stronger
limitation than the Hessian  constraint alone, and it should not
astonish that a solution obeying it may turn out unique. In particular,
for $\epsilon=1$ and $e>0$ we get a solution with $|\dot{\vec{x}}|<1$
when $\dot{\phi}=\dot{\phi}_{\pm}$, where
$$\dot{\phi}_{\pm}=\pm\,\frac{2}{\ell}\,\mu^{-1}_{\pm},\qquad
\mu_{\pm}={\sqrt{1+\br{\frac{m}{eHR}}^2}}\,\abs{1\mp \frac{2\,R}{\ell }
} -1>0.$$  The corresponding Hessian kernel vector reads
$w=\frac{\ell}{2}\,|\dot{\phi}\,|({1-R\,\epsilon\,
\dot{\phi}})\,\vec{n}\partial_{\vec{x}}\oplus\varrho\,
\dot{\phi}\,\partial_{\phi}$  with
$\varrho=\case{1-R\epsilon\dot{\phi}}{1+R\epsilon\dot{\phi}}
(1+\frac{\ell}{2}|\dot{\phi}|)$.

To understand why the frequency is fixed, suppose  that a variation
$\delta\dot{q}$ could be made collinear with $w$ for the solutions (in
which case arbitrary acceleration of the rotation frequency could be
possible). Then,
$\delta\dot{\vec{x}}=K\,\frac{\ell}{2}\,|\dot{\phi}|({1-R\,\epsilon\,
\dot{\phi}})\,\vec{n}$, $\delta{\dot{\phi}}=K\,\rho\,\dot{\phi}$ and
$\delta{\dot{\theta}}=0$ with $K$ being some function. On the other
hand, the corresponding variation of the Hessian constraint
$\vec{H}\br{\vec{n}\times\dot{\vec{x}}}=0$ gives
$\vec{H}\br{-\dot{\vec{x}}\times\delta\vec{n}+\vec{n}\times\delta\dot{\vec{x}}}=0$,
\ie, $\vec{H}\br{\vec{n}\times\delta\vec{n}}=0$ for co-rotation, or in
coordinates and evaluated at the solutions,
$|\vec{H}|\,\delta{\phi}=0$. Hence, $\delta{\phi}=0$, which means that
${\delta\dot{\phi}}=0$, a contradiction with the assumed collinearity
of $\delta{\dot{q}}$ and $w$ for $\dot{\phi}\not\equiv0$. This means
that the particular solutions must have fixed frequency, as expected.
There is also another way of seeing why the frequency is fixed. Suppose
that  the frequency has been altered at some instant for a short period
of time,  $\dot{\phi}\to\dot{\phi}+\delta\dot{\phi}$, without changing
positions and other velocities. The corresponding change in the
velocity would be
$\delta\dot{\vec{x}}=\epsilon\,R\,\vec{n}\,\delta\dot{\phi}$ and the
corresponding 'acceleration' $\delta\dot{q}$ would be proportional to
$a=\epsilon{}R\,\delta\dot{\phi}\,\vec{n}\partial_{\vec{x}}\oplus\delta\dot{\phi}\,
\partial_{\phi}$. Then, $a$ and $w$ could be collinear only when
$R=\frac{\ell}{2}$ and $\epsilon=\dot{\phi}/|\dot{\phi}|$, in which
case $\dot{\vec{x}}=\frac{\ell}{2}|\dot{\phi}|\vec{n}$ -- a condition
satisfied in the free motion  (compare with solution
\eqref{eq:gensol}). This cannot be satisfied when $|\vec{H}|\ne0$ which
is evident from the explicit solutions. When $\ell\to0$,
$\dot{\phi}_{-}$ reduces to the cyclotron frequency and when
$H\to\infty$ the frequency tends to $-R^{-1}$ -- a motion with the
velocity of light. It is important to see that $\dot{\phi}_{-}$ tends
to $0$ as $H\to0$ (or $e\to0$) in which limit one recovers the inertial
motion, not the rotational motion \eqref{eq:gensol}. Another solution
for $e>0$ has positive frequency (opposite to that expected for a
positive charge in this field) and the velocity lower than that of
light: $\dot{\phi}=\dot{\phi}_{+}$. In particular, these assumptions
are not satisfied in a region containing $\ell/2$, making $R=\ell/2$
impossible. The indeterminate frequency solution \eqref{eq:gensol} is
again not attainable in the limit $H\to0$ (or $e\to0$) even though the
condition $\dot{\vec{x}}\vec{n}=R\,|\dot{\phi}|>0$ is now satisfied.

It should be stressed that the considered motion in uniform magnetic
field is qualitatively different from the free motion
\eqref{eq:gensol}. The free model is not a simple limit of the
interacting model when $e$ tends to $0$ -- to connect both the
situations in some abstract solution manifold one has to cross the
inertial motion barrier $\dot{\phi}\equiv0$ which in a sense is
singular. For example, there is  a circular motion in magnetic field
for which $\dot{\vec{x}}\vec{n}=-\case{\ell}{2}|\dot{\phi}|<0$ when
$R=\case{\ell}{2}$,  while this inequality is impossible in free
circular  motion when
$\dot{\vec{x}}\vec{n}\equiv+\case{\ell}{2}|\dot{\phi}|>0$.

\subsection{Remarks on physical inviability of the Hessian
constraint $F_{\mu\nu}k^{\mu}\dot{x}^{\nu}=0$}

In empty space there are two qualitatively different motions possible
for the fundamental rotator: the inertial motion (with zero frequency)
and the rotational motion (with indeterminate frequency). The two
situations are qualitatively distinct. Accordingly, in the presence of
fields,  one should expect two branches of solutions that in the limit of vanishing fields pass over to two
limiting solutions: one is the inertial motion and the other is the free indeterminate rotational motion
 (we observed this in our example \ref{sec:example}).

We have seen already in section \ref{sec:example} that other models with
similar Hessian singularity do have indeterminate solutions in the
presence of external fields and that the Hessian constraint is not
sufficient to make those solutions unique at all. The same conclusion
were arrived at in \cite{bib:bratek4} where non-unique Hessian
constrained solutions were presented in the electromagnetic field for a
possible Newtonian counterpart of the fundamental rotator. This
nonuniqueness feature of singular Lagrangians is clear. Recall, that
accelerations for the internal \dof are not unique when the Hessian is
singular. The higher derivatives are thus not uniquely determinable
from the lower derivatives by the equations of motion and their
derivatives, and additional conditions are required to remove this
indeterminacy.  The  Hessian constraints (which play the role of
integrability condition of the equations) and their derivatives are not
sufficient for entire removal of this indeterminacy, except for
particular solutions with high symmetry.

It is justified to put forward a conjecture that also for the charged
fundamental rotator the solutions obeying the Hessian constrained
$F_{\mu\nu}k^{\mu}\dot{x}^{\nu}=0$  will not be unique for general
fields. This could be proved by trying to find some solutions by means
of a series expansion method and to see that for generic fields not all
expansion coefficients could be determined from the permissible initial
data. It is evident that the occurrence  of unique solutions is not in
contradiction with Hessian singularity. As we have seen in section
\ref{sec:magnetic_remark}, under certain conditions velocities cannot
be varied along the direction of the Hessian kernel vector. The
presented solutions in uniform magnetic field were found assuming
co-rotational motion. Co-rotation is a trivial solution of the Hessian
constraint, nonetheless it turned out to be very restrictive.  In
conjunction with the planar motion it allowed only for variations in
the subspace orthogonal to the kernel of the singular Hessian. In that
case the permissible variations increase the value of the action
functional so that the extremals on a constraint surface become unique.
Another possibility of showing that Hessian constrained solutions may
be non-unique would be to find an example of the indeterminate solution.
Unfortunately, the constrained equations for the fundamental rotator in
electromagnetic field are too complicated to explicitly show the branch
of non-unique solutions. Nevertheless, we could observe such solutions
in a simpler setup in section \ref{sec:example} and this should be
convincing enough.  Also in \cite{bib:gitman} a similar argumentation
for  non-uniqueness of similarly constrained solutions is presented
that can be summarized as follows:  constraints such as
\eqref{eq:constraint_vector_form} are a limitation on the possible
initial conditions, and a unique solution cannot be fixed in general, even by
selecting initial data permissible by the constraints (this statement
was also illustrated with the help of an example even simpler than that
we presented in section \ref{sec:example}). But this theoretical issue
seems only secondary in our context -- there is more important
argumentation addressing physical inviability of the charged
fundamental rotator.

Constraint \eqref{eq:constraint_vector_form} requires that for
consistency of  the equations of motion with the Hessian singularity,
the vector $\vec{n}-\dot{\vec{x}}$ must be always perpendicular to the
actual direction of the Lorentz force. The magnitude of this force is
completely irrelevant, it could be even infinitesimal. Thus, the
constraint is not dynamical, and it is not a force. The constraint is  in
conflict with the symmetries of space -- not all initial conditions
admissible for the rotator in free motion are allowed in the presence
of  electromagnetic field (for example, if initially $\dot{\vec{x}}=0$,
then the initial direction $\vec{n}$ could not be chosen arbitrary but
should be perpendicular to the electric field, even for infinitesimally
weak field). This never happens for known particles with spin for which
the initial data such as the position, orientation in space and
velocities, can be arbitrary both in the presence and absence of
external fields. The limitation on velocities and positions persists
for the charged fundamental rotator  even in the presence of the infinitely
weak field, whereas  there is nothing preferable  for positions and
directions in the configuration space \cp  and the corresponding
tangent space which both are totally symmetric.   Furthermore, in order
to satisfy constraint \eqref{eq:constraint_vector_form},  the
fundamental rotator should be instantaneously following variations in
the field irrespectively of its intensity and frequency, even for
arbitrarily large mass and spin of the rotator.  Surely, the presence
of the electromagnetic field cannot be blamed for these paradoxes -- the
motion of all other electrically charged rotators with regular
Lagrangians on \cp is not constrained in the electromagnetic field and
the initial conditions can be arbitrary, in agreement with the
symmetries of \cp (then the acceleration term is not removed from
equations such as \eqref{eq:constr1} and no constraint appears).

\section{Summary and concluding remarks}

At a given instance of time in an inertial reference frame the physical
state of a relativistic rotator  is uniquely defined by specifying
position and the associated velocities in a five-dimensional physical
configuration space identical to \cp. We found it essential that the
Lagrangian expressed in terms of the \dof on \cp should be regular,
\ie the associated Hessian should be nonsingular.

Of particular interest is considering fundamental rotators. Their
Casimir mass and spin are fixed parameters. There are two fundamental
rotators. All the other rotators are called phenomenological and their
mass is some function of the spin.

The main result of this work is to show that fundamental rotators can
be characterized by singularity of the Hessian. Despite fundamental
rotators were extensively studied in the literature this serious defect
has not been recognized. On account of this singularity the
accelerations cannot be uniquely determined from the initial state, which results in some nonuniqueness in the motion. We have found in a
covariant way the general solution and identified the indeterminate
degree of freedom. The associated velocity has the physical
interpretation of the frequency of a circular motion in the \cmf. It
can be an arbitrary function of the time -- a property characteristic of systems with gauge freedom. A physical state cannot depend on gauge
variables. To remove this paradox we conclude that fundamental rotators
interpreted as genuine rotators are defective as dynamical systems. In
contrast, the equations of motion of  phenomenological rotators are
determinate.

The Hessian singularity of fundamental rotators says that the necessary
condition for the invertibility of the equations defining momenta is
broken and  the Lagrangian description on \cp cannot be unambiguously
transformed to a Hamiltonian description with a \textit{definite}
dynamics on \cp since there are no nontrivial constraints. The  absence
of constraints follows from our construction of the Hessian null space.
We showed that the system of equations of motion on \cp is
under-determinate, only four equations are linearly independent.  This
explains the presence of the arbitrary function of the time in the general
solution.  Our analysis at the Lagrangian level has been recently
confirmed at the Hamiltonian level in \cite{bib:bratek4} where a
minimal Hamiltonian  was presented based on the Dirac formalism for the
whole family of relativistic rotators and where the Hamiltonian
equations of motion were solved both for fundamental and
phenomenological rotators. The Hamiltonian picture gives the same
indeterminate motion for fundamental rotators. A comment on this is
appropriate here. As already noted, fundamental rotators are equivalent
at the Lagrangian level to a geometric spinning particle model
suggested in \cite{bib:segal}. A  Hamiltonian formulation  presented
therein become the basis for a quantization of this system. However,
the authors did not refer to any nonuniqueness. As follows from
\cite{bib:bratek4} a unique motion at the Hamiltonian level is possible
only for phenomenological rotators. To remove this discrepancy, recall
that for a given initial data the mass and spin of phenomenological
rotators are constants of motion -- their Poisson brackets with the
Hamiltonian vanish and therefore the mass and spin can be regarded as
fixed at the Hamiltonian level and effectively put as constants into
the Hamiltonian (although one should remember that then the mass is a
function of the spin and both depend on the initial conditions). In
this sense the Hamiltonian found in \cite{bib:segal} should be
reinterpreted as that for phenomenological rotators since it cannot be
correct for fundamental rotators with indeterminate motion at the
Lagrangian level. To correctly describe a fundamental rotator at the
Hamiltonian level, an additional constraint should be included in the
Hamiltonian that would commute with other constraints of
reparametrization and projection invariance (the latter two are the
only constraints assumed in \cite{bib:segal}). Then a non-unique
solution could be obtained at the Hamiltonian level equivalent to that
found for fundamental rotators at the Lagrangian level (and this
program has been realized in \cite{bib:bratek4}). The conclusion is
that the quantum mechanics found in \cite{bib:segal} is that for
phenomenological rotators rather than that for fundamental rotators.

Our results bring up an interesting general question about the
existence of non-degenerate classical fundamental systems, that is,
without the analogous Hessian singularity for the physical \dof. This
offers us a new field of interesting investigations. The most important
in this context is to construct a comparably simple relativistic
dynamical system which would be both fundamental and non-degenerate.
Then one could find unambiguously the equivalent Hamiltonian
description with a definite dynamics and obtain its quantum mechanical
version  corresponding to the classical Lagrangian level.

The Hessian degeneracy of fundamental rotators could be removed by
introducing an interaction term with suitable nonlinearities in
velocities. However, even then there would still persist the problem of
non-uniqueness of free motion. We addressed the issue of interaction in
a more detail. We showed that an interaction term linear in the
velocities cannot remove the Hessian degeneracy, in particular this
degeneracy is not removed by the minimal coupling with the
electromagnetic field. In general, when the singularity is not removed,
Hessian constraints must appear which involve positions and velocities.
The number of such constraints equals the dimensionality of the
Hessian's null space. In particular, for the Lagrangian of fundamental
rotators on \cp a single constraint for admissible motions appears. In
the free motion this constraint is trivial showing linear dependence of
the equations of free motion. For the minimal coupling of fundamental
rotators with the electromagnetic field the constraint is nontrivial and
reads $F_{\mu\nu}k^{\mu}\dot{x}^{\nu}=0$. We gave arguments that this
constraint is physically unviable. For the similar form of interaction
no constraint is present for phenomenological rotators and their motion
is unique.

It might be expected that the presence of a constraint could make
solutions unique. This is not the case. In a simple example of a point
particle with singular Hessian we illustrated that there can be
solutions which are still non-unique. Occasionally, unique motions are
nevertheless possible in external fields. We illustrated the mechanism
of appearance of such unique solutions by considering a co-rotational
motion of the fundamental relativistic rotator in the uniform magnetic
field.

It is worth mentioning that the fundamental relativistic rotator is not
an isolated example when imposing fundamental conditions goes hand by
hand with the Hessian singularity. Recently, an extended class of
rotators was studied in which fundamental systems also proved
similarly defective \cite{bib:bratek2}. One of the systems turned out
equivalent to a universal spinning particle model
considered many years ago by Lyakhovich, Segal and Sharapov
\cite{bib:segal2}.

\section*{Acknowledgements}
I would like to acknowledge Professor Andrzej Staruszkiewicz for his
always invaluable and stimulating discussions.

\appendix

\section{\label{app:fr}Solution of fundamental conditions}

The momenta canonically conjugated to $x$ and $k$ for the action
functional \eqref{eq:act_f} are, respectively,
\begin{equation}\fl\label{eq:momenta}P\equiv-\frac{\partial{}L}{\partial{}\dot{x}}=
\frac{mf(Q)}{\sqrt{\dd{x}}}\dot{x}
-2m\,Qf'(Q)\frac{\sqrt{\dd{x}}}{k\dot{x}}k, \quad
\Pi\equiv-\frac{\partial{}L}{\partial{}\dot{k}}=2m\,Qf'(Q)\frac{\sqrt{\dd{x}}}{\dd{k}}
\dot{k}.\end{equation} The invariance of the Hamilton's action with
respect to Poincar{\'e} transformations implies that  the momentum
vector $P$ and the angular momentum tensor $M$ are conserved for
solutions. Indeed, since the general variation of the Lagrangian reads
$$\delta{L}=-\frac{d}{d\tau}\br{P\delta{x}+\Pi\delta{{k}}}+
\dot{P}\delta{x}+\br{\dot{\Pi}+2m\,Qf'(Q)\frac{\sqrt{\dd{x}}}{k\dot{x}}\dot{x}}\delta{k},
$$  then, for infinitesimal global space-time translations $\epsilon$
and rotations $\Omega$ of solutions, one has the following implications
$\br{\delta{}x=\epsilon=\const,\,\delta{k}=0}\Rightarrow P=\const$ and
$(\delta{x}^{\mu}=\Omega^{\mu}_{\phantom{\mu}\nu}x^{\nu},\,
\delta{k}^{\mu}=\Omega^{\mu}_{\phantom{\mu}\nu}k^{\nu},\,
\Omega_{\mu\nu}=\const,\,{\Omega_{(\mu\nu)}=0})\,\Rightarrow\,
M_{\mu\nu}\equiv{}x_{\mu}P_{\nu}-x_{\nu}P_{\mu}+k_{\mu}\Pi_{\nu}-k_{\nu}\Pi_{\mu}=\const
$ For the purpose of this section it suffices  to keep in mind that
$kk=0$, however, in order to find the equations of motion in a
covariant form, one may add to the Hamilton's action an appropriate
term with a Lagrange multiplier (this is done in \ref{sec:solution}).
The Casimir invariants of the Poincar\'{e} group are $PP$ and $WW$,
where $W$ is the Pauli-Luba\'{n}ski (space-like) pseudo-vector defined
in \eqref{eq:fundam_condit}; hence, \begin{eqnarray*} \fl\qquad\quad
PP=m^2\br{f^2(Q)-4Qf(Q)f'(Q)},\\ \fl \qquad\quad
WW=-4m^4\frac{Q^2f^2(Q)f'^2(Q)}{(\dd{k})^2} \left|\begin{array}{ccc}
kk& k\dot{k}& k\dot{x}\\ \dot{k}k& \dot{k}\dot{k}& \dot{k}\dot{x}\\
\dot{x}k& \dot{x}\dot{k}& \dot{x}\dot{x}\\ \end{array}\right|
=-4m^4\ell^2Qf^2(Q)f'^2(Q).\end{eqnarray*} Now, we find all
relativistic rotators satisfying fundamental conditions
\eqref{eq:fundam_condit}. It follows from $PP\equiv{}m^2$ that
$f(Q)=\sqrt{1\pm a^2\sqrt{Q}}$, $a>0$. Condition
$WW=-\frac{1}{4}m^4\ell^2$ will be satisfied for $a=1$.  This gives
action \eqref{eq:action}. A similar derivation of this result was
carried out independently in \cite{bib:schaefer}
(cf \cite{bib:bratek1}).

\section{\label{sec:hessian_calculation} Calculation of the Hessian
determinant and construction of the null space of the singular
Hessian.}

Prior to calculation of the Hessian determinant  associated with the
\dof \cp for action \eqref{eq:act_f}, the auxiliary \dof must be
eliminated which are irrelevant to the dynamics. It suffices to
consider some particular map adapted to the configuration space \cp.
The generality of the calculation is not lost by choosing such a map,
since we are interested in the universal dependence of the Hessian
determinant on function $f(Q)$ and its derivatives. In regard to the five
physical \dof, it is convenient to use the (dimensionless) Cartesian
coordinates $\br{X^1,X^2,X^3}$ for the position vector, and the
spherical angles $(\vartheta,\varphi)$ for the null direction, whereas
the gauge is fixed such that $\tau\equiv{}\ell{}{u}$ and  $k^0\equiv1$,
where $\ell {u}$ is the time coordinate in the Minkowski spacetime.
Specifically, \begin{eqnarray*} &x^0({u})=\ell{}{u},\quad
&\vec{x}({u})=\ell\sq{X^1({u}),X^2({u}),X^3({u})},\\ &k^0({u})=1, \quad
&\vec{k}({u})=\sq{\sin{\vartheta({u})}\cos{\varphi({u})},
\sin{\vartheta({u})}\sin{\varphi({u})},\cos{\vartheta({u})}}.
\end{eqnarray*} In this parametrization the Lagrangian in action
\eqref{eq:act_f} is proportional to function $\mathcal{L}$
$$\mathcal{L}\br{V,W}=\sqrt{1-V^TV}f(Q),\qquad
Q=\frac{W^TW}{\br{1-N^TV}^2},$$ where
$$V=\left(\begin{array}{c}{X}^1{}'({u})\\ {X}^2{}'({u})\\
{X}^3{}'({u})\end{array}\right), \qquad
W=\left(\begin{array}{c}{\vartheta}'({u})\\
{\varphi}'({u})\sin{\vartheta({u})}\end{array}\right), \qquad
N=\left(\begin{array}{c}\sin{\vartheta({u})}\cos{\varphi({u})}\\
\sin{\vartheta({u})}\sin{\varphi({u})}\\
\cos{\vartheta({u})}\end{array}\right).$$ The Hessian determinant
calculated with respect to the velocities ${X}^1{}'({u})$,
${X}^2{}'({u})$, ${X}^3{}'({u})$, ${\vartheta}'({u})$ and
${\phi}'({u})$, is proportional to the determinant of the following
symmetric matrix of size $5\times5$
\begin{equation}\label{eq:hessian_matrix}H=\left[\begin{array}{cc}A&B\\
B^T&C\end{array}\right],\end{equation} where\footnote[1]{ Note that
$A=A^T$ and $C=C^T$ have size $2\times{2}$ and $3\times{3}$, but $B$
and $B^T$ are matrices of different shape, of size $2\times{3}$ and
$3\times{2}$, respectively. Note also the obvious thing that the order
of multiplication is important, e.g $WV^T$ is a rectangular matrix with
$2$ rows and $3$ columns, $NV^T$ is a $3\times3$ square matrix, whereas
$N^TV=V^TN$ is a scalar product of column vectors $N$ and $V$. }
\begin{small} \begin{eqnarray*}\fl A=
2Qf'(Q)\frac{\sqrt{1-V^TV}}{W^T{W}}\br{I+ 2\frac{Q
f''(Q)}{f'(Q)}\frac{WW^T}{W^T{W}}} \\ \fl B=
2Qf'(Q)\frac{\sqrt{1-V^TV}}{W^T{W}} \br{2\sq{1+\frac{Q
f''(Q)}{f'(Q)}}\frac{WN^T}{1-N^T{V}}- \frac{WV^T}{1-V^T{V}} }, \\ \fl
C= -\frac{f(Q)}{\sqrt{1-V^TV}}\cdot
\left(I+\frac{VV^T}{1-V^TV}
+2\frac{Q\,f'(Q)}{f(Q)}\right.\\ \left.\times\sq{\frac{NV^T+VN^T}{1-N^T{V}}-\br{3+2\frac{Q
f''(Q)}{f'(Q)}} \frac{1-V^T{V}}{\br{1-N^T{V}}^2}NN^T}\right).
\end{eqnarray*}\end{small}

\noindent
By $I$ are denoted the identity matrices of appropriate size. The
elements of matrices $A$, $B$ and $C$ are numerically equal to the
respective second derivatives
$$A^{i'}_{j'}\hat{=}\frac{\partial^2\mathcal{L}}{\partial{W^{i'}}\partial{W^{j'}}},
\qquad
B^{i'}_{j}\hat{=}{\frac{\partial^2\mathcal{L}}{\partial{W^{i'}}\partial{V^{j}}}}
={\frac{\partial^2\mathcal{L}}{\partial{V^{j}}\partial{W^{i'}}}}\hat{=}\br{B^{T}}^{j}_{i'},
\qquad
C^{i}_{j}\hat{=}\frac{\partial^2\mathcal{L}}{\partial{V^{i}}\partial{V^{j}}},$$
$i,j=1,2,3$, $i',j'=1,2$. Owing to the block structure of matrix $H$,
calculation of its determinant can be significantly simplified. First,
one employs the following identity: $$\left[\begin{array}{cc}A&B\\
B^T&C\end{array}\right]=\left[\begin{array}{cc}A&0\\
B^T&I\end{array}\right]\cdot\left[\begin{array}{cc}I&A^{-1}B\\
0&C-B^TA^{-1}B\end{array}\right],$$ holding for a block matrix composed
of  matrices of mutually compatible dimensions. Hence, one concludes
that $\det(H)=\det(A)\det(C-B^TA^{-1}B)$. By applying the Sylvester
determinant theorem\footnote[2]{In general, Sylvester's  theorem states
that $\det(I_{{m\times{}m}} + RS) = \det(I_{{n\times{}n}} + SR)$ for
matrices $R$ and $S$ of size $m\times{}n$ and $n\times{}m$,
respectively, where $I_{{m\times{}m}}$ and $I_{{n\times{}n}}$ are unit
matrices. In particular, it follows that for column vectors $a$ and $b$
of size $n$ and a nonsingular matrix $M$ of size $n\times{}n$ one has
$\det(M + ab^T)=\det\br{M\br{I_n + M^{-1}ab^T}}
=\br{\det{M}}\det\br{I_1 + b^TM^{-1}a}=\br{\det{M}}\br{1+b^TM^{-1}a} $.
} one can easily calculate $\det(A)$
$$\br{2Qf'(Q)\frac{\sqrt{1-V^TV}}{W^T{W}}}^{-2}\det(A)=\det\br{I+
2\frac{Qf''(Q)}{f'(Q)}\frac{WW^T}{W^T{W}}} =1+2\frac{Qf''(Q)}{f'(Q)}.
$$ The inverse of $A$ can also be easily found by supposing that
$A^{-1}=a\br{I+b WW^T}$ with $a$ and $b$ to be determined from the
condition $A^{-1}A=I=AA^{-1}$. The result is
$$A^{-1} =\frac{W^T{W}}{2Qf'(Q)\sqrt{1-V^TV}}\br{I-
2\frac{Qf''(Q)}{f'(Q)+2Qf''(Q)}\frac{WW^T}{W^T{W}}}.$$ By noting that
$\br{XW^T}\br{WW^T}\br{WY^T}=\br{W^TW}^2\br{XY^T}$, etc., holds for
column vectors $W,X,Y$, one finds  that \begin{eqnarray*}\fl
C-B^TA^{-1}B=-\frac{f(Q)}{\sqrt{1-V^TV}} \left[
\br{I+\frac{VV^T}{1-V^TV}}+\dots \right. \\ \fl\quad  \left.
\dots\frac{2Qf'(Q)}{f(Q)\br{1+\frac{2Qf''(Q)}{f'(Q)}}}
\frac{1-V^TV}{\br{1-N^TV}^2} \br{N-\frac{1-N^TV}{1-V^TV}V}
\br{N^T-\frac{1-N^TV}{1-V^TV}V^T}\right]. \end{eqnarray*}This is again
a square matrix to which the Sylvester determinant theorem applies
\begin{small} \begin{eqnarray*}\fl
-\br{\frac{f(Q)}{\sqrt{1-V^TV}}}^{-3}\det{\br{C-B^TA^{-1}B}}=
\det\br{I+\frac{VV^T}{1-V^TV}} \left[ 1+
\frac{2Qf'(Q)}{f(Q)\br{1+\frac{2Qf''(Q)}{f'(Q)}}}\,\cdot\right.\\
\left. \fl\qquad\dots\cdot\frac{1-V^TV}{\br{1-N^TV}^2}
\br{N^T-\frac{1-N^TV}{1-V^TV}V^T} \br{I-VV^T}
\br{N-\frac{1-N^TV}{1-V^TV}V}\right]=\\
=\frac{1}{1-V^TV}\br{1+\frac{2Qf'(Q)}{f(Q)\br{1+2\frac{Q
f''(Q)}{f'(Q)}}}}. \end{eqnarray*} \end{small}

\noindent
Finally, on expressing $W^TW$ by $Q$ in the  formula for $\det{A}$
derived earlier, it follows that \begin{equation}
\fl\label{eq:hessian_appendix}\det{H}=-\frac{4f(Q)^3f'(Q)^2}{
\br{1-N^TV}^4\br{1-V^TV}^{3/2}}\br{1+2Q\br{
\frac{f'(Q)}{f(Q)}+\frac{f''(Q)}{f'(Q)}}}.\end{equation}

\subsection{\label{sec:null_space_calc}The null space of the Hessian of
the fundamental relativistic rotator}

Let denote a $5$-component kernel vector from this space by $w$. Vector
$w$ can be regarded as a direct sum of two vectors of dimension $2$
(associated with the angular variables) and of dimension $3$
(associated with the position variables). Since matrices $A$, $B$ and
$C$  were constructed using vectors $W$, $N$ and $V$, the vector $w$
can be supposed to be represented as a column vector, $w=[\alpha W^T,
\beta N^T+\gamma V^T]^T$, with unknowns $\alpha$, $\beta$, $\gamma$.
This is an ansatz for the vector $w$. A kernel vector of the Hessian
matrix \eqref{eq:hessian_matrix} must satisfy two vector equations
$\alpha{}AW+\beta{}BN+\gamma{}BV=0$ and
$\alpha{}B^TW+\beta{}CN+\gamma{}CV=0$. On taking scalar products with
$W$, $N$ and $V$, one obtains three scalar equations for unknowns $\alpha$,
$\beta$, $\gamma$: $$\left[\begin{array}{ccc} W^TAW & W^TBN&W^TBV\\
N^TB^TW&N^TCN&N^TCV\\ V^TB^TW&V^TCN&V^TCV \end{array}\right]\cdot
\left[\begin{array}{c}\alpha\\ \beta\\ \gamma\end{array}\right]
=\left[\begin{array}{c}0\\0\\0\end{array}\right].$$ The determinant of
the square matrix  reads $$2\frac{V^TV-\br{N^TV}^2} {{\left( 1 - {V^TV}
\right) }^{\frac{3}{2}}} \,Q\,f'(Q)\,{f(Q)}^2 \br{1+2Q\br{
\frac{f'(Q)}{f(Q)}+\frac{f''(Q)}{f'(Q)}}}$$ and is proportional to the
Hessian determinant \eqref{eq:hessian_appendix}, in accord with the
expectation that a nontrivial kernel vector exists only when the
Hessian matrix is singular. To the extent of an arbitrary numerical
factor, there is only one nontrivial kernel vector ($f'(Q)\ne0$ by assumption):\begin{equation}\label{eq:nullvector}\fl\phantom{XXX}
w=\frac{2(1-N^TV)^2+(1-2N^TV+V^TV)\sqrt{W^TW}}{1-V^TV}\frac{W}{\sqrt{W^TW}}\oplus\br{N-V}.
\end{equation}

\section{\label{sec:solution} Construction of solutions for the
fundamental relativistic rotator}

In this section we find a covariant form of solutions to the equations
of motion resulting from the Hamilton action \eqref{eq:action}.  For
concreteness we chose the Lagrangian with the $+$ sign (for the other
Lagrangian the solution can be found in a similar manner). The
generalized momentum $p^{\mu}$ corresponding to spacetime coordinates
${x}^{\mu}$ is \begin{eqnarray}\label{eq:p}
{p}_{\mu}\equiv-\frac{\partial{L}}{\partial{\dot{x}^{\mu}}}=m\br{\exp{\Psi}
u_{\mu}-\sinh\br{\Psi}\frac{k_{\mu}}{ku}},\nonumber\\ \fl
\mathrm{where}\\
u^{\mu}\equiv\frac{\dot{x}^{\mu}}{\sqrt{\dd{x}}},\qquad
\exp{2\Psi}\equiv{\sqrt{- \ell^2\frac{\dd{k}}{\br{\dot{x}k}^2}}+1},
\quad (\Psi\geqslant0).\nonumber\end{eqnarray} It follows that
$pp=m^2$. The generalized momentum corresponding to the null direction
${k}^{\mu}$ is
$$\pi_{\mu}\equiv-\frac{\partial{L}}{\partial{\dot{k}^{\mu}}}=m\,\sqrt{\dd{x}}\,
\sinh\br{\Psi}\frac{\dot{k}_{\mu}}{{\dd{k}}}=-\frac{m^2\ell}{2
pk}\,\frac{\dot{k}_{\mu}}{\sqrt{-\dd{k}}},$$ where  the identity
$2pk\sqrt{\dd{x}}\sinh{\Psi}=\ell{}m\sqrt{-\dd{k}}$ resulting from
(\ref{eq:p}) has been used. A convenient way of deriving the equation
of motion for $k^{\mu}$, without the need of introducing the internal
coordinates on the cone $kk=0$,  is to find a conditional extremum of
functional (\ref{eq:action}) subject to the condition $kk=0$. This is a
standard variational problem with subsidiary conditions
\cite{bib:hilbert} -- the stationary value of functional
(\ref{eq:action}) with the condition $kk=0$, can be found by
supplementing the functional with an additional term
$\int\ud{\tau}(-)\Lambda(\tau)kk$, containing a Lagrange multiplier
$\Lambda(\tau)$.  By varying such extended action with respect to
$\Lambda$, one restores the condition $kk=0$, whereas the variation
with respect to vector $k^{\mu}$, yields the following equation:
$\dot{\pi}_{\mu}+\partial_{{k}^{\mu}}L-2\Lambda k_{\mu}=0$. By
contracting it with vector $p^{\mu}$, one finds the unknown function
$\Lambda(\tau)$, and hence, the equation of motion for $k$
\begin{equation}\label{eq:tensor}\br{\dot{\pi}_{\nu}+
\frac{\partial{}L}{\partial{k}^{\nu}}}\br{\delta^{\nu}_{\phantom{\nu}\mu}-
\frac{p^{\nu}k_{\mu}}{pk}}=0,\qquad  kk=0.\end{equation} This  equation
can be recast in a form having a very clear geometrical meaning.
Firstly, a null vector $k^{\mu}$ can always be written as
$k^{\mu}=h\br{m^{-1}p^{\mu}+n^{\mu}}$, where $n^{\mu}$ is a unit
space-like vector orthogonal to timelike vector $p^{\mu}$, and
$h=m^{-1}p^{\mu}k_{\mu}$. Secondly, for describing a space-like curve
$n^{\mu}(\tau)$, it is more natural to regard its arc length
$$\phi\br{\tau}= \int\ud{\tau}\sqrt{-\dd{n}},\qquad nn=-1,\quad np=0,
$$ as the independent variable, rather than any other. Furthermore, the
momentum $p^{\mu}$ is conserved, $p^{\mu}(\tau)=P^{\mu}$, where
$P^{\mu}$ is a constant vector such that $PP=m^2$. Now, making use of
these observations,  equation (\ref{eq:tensor}) can be reduced (up to
unimportant $h$-dependent factor) to the following equation for
$n^{\mu}$:
\begin{equation}\label{eq:n}\frac{\ud{}^2n^{\mu}}{\ud{}\phi^2}+n^{\mu}=0,
\qquad nn=-1,\quad nP=0.\end{equation} This is nothing but the equation
for great circles on a unit sphere in the subspace orthogonal to
$P^{\mu}$ (then $\phi(\tau)$ is the angle). Expressed in terms of
$n^{\mu}(\phi)$, the Pauli-Luba\'{n}ski spin-vector reads
$$W^{\mu}=\frac{1}{2}m\ell\epsilon^{\mu\alpha\beta\gamma}n_{\alpha}
\frac{\ud{n_{\beta}}}{\ud{\phi}}P_{\gamma}.$$ This constant
pseudo-vector is orthogonal to the plane spanned by $n^{\mu}$ and
$\frac{\ud{n^{\mu}}}{\ud{\phi}}$; thus, together with $P^{\mu}$,  it
can be used  to construct solutions.

A parametric description of a specific circle from the family of
solutions can be visualized as a continuous action of an elliptic
Lorentz transformation upon some fixed unit spatial vector $N^{\mu}$
orthogonal to $W^{\mu}$ and $P^{\mu}$. Such a transformation must leave
invariant two null directions
$K_{\pm}^{\mu}=\frac{1}{\sqrt{2}}\br{\frac{P^{\mu}}{m}\pm\frac{W^{\mu}}{
\frac{1}{2}m^2\ell}}$. Parameterized by the elliptic angle $\phi$, the
general solution for $n^{\mu}(\phi)$ is thus easily found to be
$$n^{\mu}\br{\phi}
=N^{\mu}\cos{\phi}-\frac{\epsilon^{\mu\nu\alpha\beta}N_{\nu}W_{\alpha}P_{\beta}}{
\frac{1}{2}m^3\ell}\sin{\phi}, \qquad NN=-1,\quad NW=0,\quad NP=0.$$
This is indeed the general solution to equation (\ref{eq:n}). As was to
be anticipated from the independence of Hamilton's action
(\ref{eq:action}) upon scaling of the null vector $k$ by arbitrary
function, there is no constraint imposed on function $h$   by the
equations of motion, thus, without loss of generality, one may set
$h\equiv1$. Finally, the corresponding null direction and spacetime
position   can be found from \begin{equation}\label{eq:sol}\fl
k^{\mu}={n^{\mu}\br{\phi{\br{\tau}}}+\frac{P^{\mu}}{m}},\qquad
\frac{{\dot{x}}^{\mu}}{\sqrt{\dd{x}}}=\frac{P^{\mu}}{m}\cdot\cosh{\Psi\br{\tau}}+
n^{\mu}\br{\phi{\br{\tau}}}\cdot\sinh{\Psi\br{\tau}}.\end{equation} The
second equation in \eqref{eq:sol} comes from the Noether integral
(\ref{eq:p}). Since the parameter $\tau$ is arbitrary, one may fix it
by requiring that $\dot{n}\dot{n}=-1$, in which case
$\phi(\tau)\equiv\tau$, or alternatively that $P\dot{x}\equiv{m}$, in
which case $\tau$ would be the proper time in the \cmf.

Function $\Psi$ is a reparametrization invariant Lorentz scalar,
\textit{e.g.} $\tanh\br{\Psi\br{\tau}}=-m\frac{n\dot{x}}{P\dot{x}}$. It
can equally well  be regarded as a function of $\phi$ by fixing the
arbitrary parameter $\tau$, so as $\tau\equiv\phi$. The equations of
motion, expressed in terms of the independent variable $\phi$, are
satisfied by any $\Psi(\phi)$. Then, as is seen from (\ref{eq:sol}),
$k^{\mu}$ is a definite function of the angle $\phi$,  whereas the
4-velocity $u^{\mu}=\frac{\dot{x}^{\mu}}{\sqrt{\dot{x}\dot{x}}}$ is
not: $un=-\sinh\br{\Psi\br{\phi}}$. This cannot happen for a well-behaved dynamical system, in which situation, after fixing $\tau$ the
same way,  one would expect to obtain a definite function $\Psi(\phi)$,
being constant for a uniform rotation.

It is rather upsetting to find out that, after fixing the
parametrization, say by requiring that $P\dot{x}\equiv{m}$, solution
(\ref{eq:sol}) remains completely undetermined  by the initial
conditions. This observation was the main motivation for this paper as
it shows that the fundamental relativistic rotator is defective as a
dynamical system. The reason for this arbitrariness  is explained in
section \ref{sec:cauchy}.

The auxiliary function $\Psi$ defined in (\ref{eq:p}) allows not only
for concise notation, but it also has a definite meaning. Namely,
$\Psi$ is the hyperbolic angle between the momentum $p^{\mu}$ and the
world-velocity $u^{\mu}$, $pu=m\cosh{\Psi}$. It is related to the time
dependence of rotation of the null direction in the \cmf. The proper
time in this frame increases by
$\ud{t}=\br{m^{-1}P_{\mu}}\dot{x}^{\mu}\ud{\tau}$ with every
infinitesimal displacement $\ud{x}^{\mu}=\dot{x}^{\mu}\ud{\tau}$ of the
rotator. It follows from equation (\ref{eq:sol})  that
$\sqrt{\dot{x}\dot{x}}\,\sinh{\Psi}=-n_{\mu}\dot{x}^{\mu}$ and, in
conjunction with the definition of $\Psi$ in equation (\ref{eq:p}),
$2\sqrt{\dot{x}\dot{x}}\,\sinh{\Psi}\ud{\tau}${}$=\ell\sqrt{-\dot{n}\dot{n}}\,
\ud{\tau}${}$\equiv\ell\,|\dot{\phi}|\,\ud{\tau}$, where
$\dot{\phi}\,\ud{\tau}$ is the corresponding change in the angular
position of the null direction,  observed in this frame. Hence, the
angular speed of the null direction in the \cmf is
$$\abs{\frac{\ud{\phi}}{\ud{t}}}=
-\frac{2}{\ell}\cdot\frac{n_{\mu}u^{\mu}}{\frac{P_{\mu}}{m}u^{\mu}}=\frac{2}{\ell}
\tanh{\Psi} <\frac{2}{\ell}$$ in agrement with the relation between the
frequency and the rotational speed ($\tanh{\Psi}$) on a circular orbit
of radius $\ell/2$.

To solve equation (\ref{eq:sol}) for $x$, we choose the arbitrary
parameter $\tau$ so that $\tau\equiv{}t$ (hereafter
$\dot{{}}\equiv\frac{\ud}{\ud{t}}$). Now, on account of the earlier
definition of $t$, $\ud{t}=\br{m^{-1}P_{\mu}}\dot{x}^{\mu}\ud{\tau}$,
one has $P\dot{x}\equiv{}m$, or equivalently,
$\cosh{\Psi}=\br{\dd{x}}^{-1/2}$. Hence,
$\dot{x}^{\mu}\ud{t}=\frac{P^{\mu}}{m}\ud{t}+n^{\mu}\tanh{\Psi}\,\ud{t}=
\frac{P^{\mu}}{m}\ud{t}+\frac{\ell}{2}n^{\mu}\ud{\phi}$, and finally,
integration gives $x^{\mu}(t)$. To show the solution in explicit form
it is best to use a vector $r^{\mu}$ defined by
$\dot{r}^{\mu}(t)=n^{\mu}(\phi(t))\dot{\phi}(t)$ rather than $n^{\mu}$,
then $|{\dot{\phi}(t)}|=\sqrt{-\dot{r}(t)\dot{r}(t)}$ (cf section
\ref{sec:gensol} for the final result).

\centering{
\rule{0.190983\textwidth}{1pt}%
\rule[-0.5pt]{0.118034\textwidth}{2pt}%
\rule[-1pt]{0.381966\textwidth}{3pt}%
\rule[-0.5pt]{0.118034\textwidth}{2pt}%
\rule{0.190983\textwidth}{1pt}}

\end{document}